# Dual-Attention Enhanced BDense-UNet for Liver Lesion Segmentation

Wenming Cao[1], Philip L.H. Yu[1], Gilbert C.S. Lui[1], Keith W.H. Chiu[2], Ho-Ming Cheng[3], Yanwen Fang[1], Man-Fung Yuen[3,4], Wai-Kay Seto[3,4,5]

[1] Department of Statistics and Actuarial Science, The University of Hong Kong
[2] Department of Diagnostic Radiology, The University of Hong Kong
[3] Department of Medicine, The University of Hong Kong
[4] State Key Laboratory for Liver Research, The University of Hong Kong
[5] Department of Medicine, The University of Hong Kong-Shenzhen Hospital, Shenzhen, China

**Abstract.** In this work, we propose a new segmentation network by integrating DenseUNet and bidirectional LSTM together with attention mechanism, termed as DA-BDense-UNet. DenseUNet allows learning enough diverse features and enhancing the representative power of networks by regulating the information flow. Bidirectional LSTM is responsible to explore the relationships between the encoded features and the up-sampled features in the encoding and decoding paths. Meanwhile, we introduce attention gates (AG) into DenseUNet to diminish responses of unrelated background regions and magnify responses of salient regions progressively. Besides, the attention in bidirectional LSTM takes into account the contribution differences of the encoded features and the up-sampled features in segmentation improvement, which can in turn adjust proper weights for these two kinds of features. We conduct experiments on liver CT image data sets collected from multiple hospitals by comparing with state-of-the-art segmentation models. Experimental results indicate that our proposed method DA-BDense-UNet has achieved comparative performance in terms of dice coefficient, which demonstrates its effectiveness.

## 1    Introduction

Hepatocellular carcinoma (HCC)HCC is one of common cancers in the Asia-Pacific region and in 2018 accounted for 10.3% of cancer death in Hong Kong[1]. Early diagnosis and detection of HCC can help to improve the medical treatment. Typically, as recommended by international guidelines, patients with suspicious HCC are diagnosed radiologically by using either computerized tomography (CT) or magnetic resonance (MR) scans to provide the detailed information and improve the accuracy of diagnosis without biopsy. Artificial intelligence has emerged as an important technological advancement in medical diagnostics. Traditionally, slices of CT scan images are interpreted by clinicians visually and the accuracy of diagnosis heavily relies on the experience of clinicians. Thus, accurate diagnosis of liver lesions could be a challenging task, and if serial scans are required, there can be a significant time lag till diagnosis, potentially delaying curative treatment. . Recently, researches adopt deep-learning methods to diagnose HCC using multi-phase CT images. Sun *et al*. [3] design a multi-channel fully convolutional network to segment tumors from multi-phase contrast-enhanced CT images, where one network for each phase of CT images is trained and high-level features of multi-phase images are fused for tumor segmentation. Todoroki *et al*. [4] adopt deep convolutional neural network for liver tumor classification based on liver segmentation [9, 10], where contrast-enhanced (multi-phase) CT images of hepatic tumors are investigated. Lee *et al*. [6] propose an optimized version of single shot multibox detector (SSD) [11], which groups convolutions for multi-phase features and leverages information of multi-phase CT images. Liang *et al*. [8] classify multi-phase CT images of focal liver lesions by combining convolutional networks and recurrent networks. Ouhmichi *et al*. [5] propose a cascaded convolutional neural network based on U-Net and design two strategies to fuse multiphase information: 1) concatenating in a multi-dimensional features map on input layer, *i.e.*, dimensional multiphase (DMP) and 2) output maps are computed independently for each phase before being merged to produce the final segmentation, *i.e.*, multiphase fusion (MPF).

---

[1] https://www.fhb.gov.hk/download/press_and_publications/otherinfo/190700_hkcs/e_hkcs_fully.pdf

Although the above works have achieved satisfactory segmentation performances, they do not consider relationships between the encoded features in the encoding path and the up-sampled features in the decoding path in the UNet. Besides, these methods tend to rely on an additional object detection model to localize tumor position for a reliable segmentation. To address these issues, this paper proposes an enhanced segmentation model that combines DenseUNet and bidirectional long short term memory together with two types of attention mechanisms into these modules, which is referred to as DA-BDense-UNet. The advantages of this model are as follows:

1) DenseUNet is capable of learning enough diverse informative features and enrich the expressive capability of networks by regulating the information flow over the network;

2) Bidirectional LSTM explores relationships between the encoded features and the up-sampled features, and provides a new strategy of feature fusion more than feature concatenation;

3) The attention gate in DenseUNet restrains response of feature maps corresponding to unrelated background regions and highlights response of salient lesion regions progressively;

4) The attention mechanism in bidirectional LSTM takes into account the contribution differences of both the encoded features and the decoded features and identifies the importance of feature maps from the perspective of channels.

## 2 The Proposed Methodology

In this section, we describe our proposed model DA-BDense-UNet for liver CT image segmentation task in detail. The overall framework of our DA-BDense-UNet is shown in Fig. 1, which includes three modules: Dense-UNet, BD-LSTM and Dual Attention.

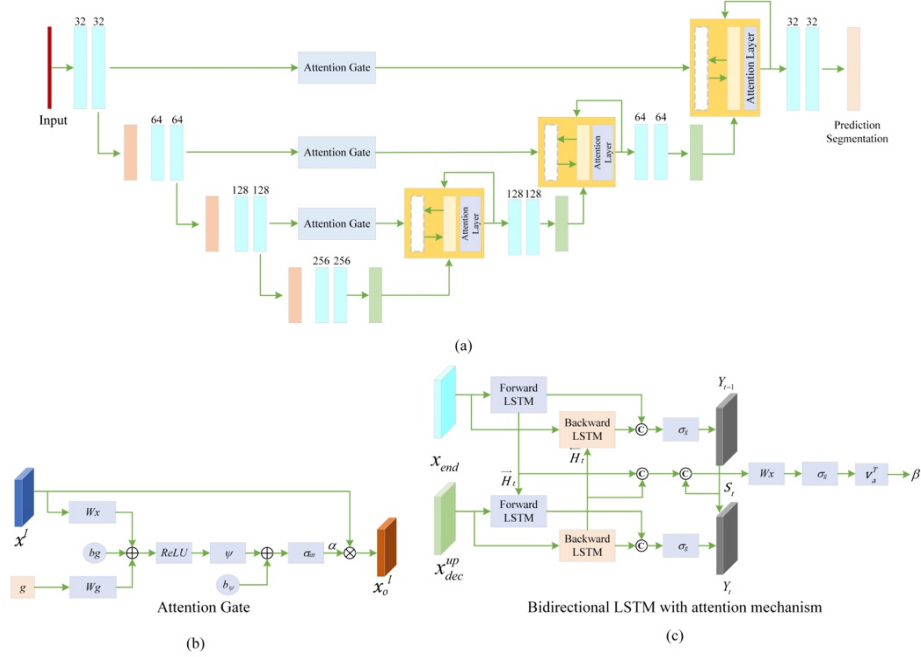

Fig. 1 The overall framework of our proposed DA-BDense-UNet.

### 2.1 Dense-UNet module

Inspired by the success of UNet and densely connected convolution network (Dense-Net), Dense-UNet has been proposed by changing the backbone of UNet from fully connected convolutional networks (FCN) to DenseNet. DenseNet enjoys the following advantages: 1) it encourages the network to learn enough diverse feature candidates rather than redundant ones; 2) it enhances the learning capability of the network through effectively allowing the information flow and reusing features; 3) it reduces the risk of gradient exploding or vanishing, and the gradients are quickly sent to their respective expected locations in the backward propagation path.

### 2.2 BD-LSTM module

Bi-Directional Long Short Term Memory (BD-LSTM) has been widely used to learn the temporal relationship of data in natural language processing. In our approach, we shall explore correlations between the encoded features in the encoder and the corresponding up-sampled features in the decoder in two directions. Basically, a BD-LSTM unit is composed of three gates and a cell. These three gates are an input gate, a forget gate and an output gate respectively, and are responsible to control the amount of information into and out of the cell. The cell aims to remember values over temporal dimension. BD-LSTM can be formulated as follows:

$$i_t = \sigma_g(W_f x_t + U_f h_{t-1} + b_f)$$
$$f_t = \sigma_g(W_i x_t + U_i h_{t-1} + b_i)$$
$$o_t = \sigma_g(W_o x_t + U_o h_{t-1} + b_o)$$
$$c_t = f_t \circ c_{t-1} + i_t \circ \sigma_c(W_c x_t + U_t h_{t-1} + b_c)$$
$$h_t = o_t \circ \sigma_h(c_t)$$

where $\circ$ denotes the element-wise product. The subscript $t$ denotes the $t^{th}$ time stamp. $\sigma_g$ is the sigmoid function, $\sigma_c$ is the hyperbolic tangent function, and $\sigma_h(x) = x$.

The output of BD-LSTM by considering both the forward and backward hidden states can be defined by:

$$Y_t = \sigma_g\left(W_y^{\vec{H}} \vec{H}_t + W_y^{\overleftarrow{H}} \overleftarrow{H}_t + b_y\right).$$

### 2.3 Dual Attention Mechanism

There are two kinds of attention mechanisms in our method, working in Dense-UNet and BD-LSTM module, respectively. Specifically, in standard CNNs like the encoder of Dense-UNet, the receptive field becomes larger for capturing the contextual information when the depth of convolution layers increases. In other words, features learnt by convolution kernels construct relationships between objects in a coarser grid level. However, it is still challenging to avoid the risk of false-positive predictions for small objects with larger shape variability. Existing segmentation approaches introduce an auxiliary object localization subnetwork to address this issue, which results in a large amount of network parameters. Practically, the object localization subnetwork can be replaced by the attention gates (AGs) capable of pruning responses of unrelated contextual regions and identifying related salient regions layer by layer. For each pixel vector $x_i^l \in \mathbb{R}^{F_l}$ where $F_l$ denotes the number of convolution kernels in the $l^{th}$ layer, the output of AGs is the element-wise product of feature responses and the corresponding attention activation responses, which is defined by:

$$\hat{x}_i^l = x_i^l \alpha_i^l,$$

where

$$\alpha_i^l = \sigma_{att}\left(q_{att}^l(x_i^l, g_i; \Theta_{att})\right),$$
$$\text{and } q_{att}^l = \psi^T\left(\sigma_1(W_x^T x_i^l + W_g^T g_i + b_g)\right) + b_\psi,$$

where $\sigma_1$ and $\sigma_{att}$ are the ReLU and the sigmoid function, respectively. AG is parametrized by $\Theta_{att}$ including $W_x \in \mathbb{R}^{F_l \times F_a}$, $W_g \in \mathbb{R}^{F_g \times F_a}$, $\psi \in \mathbb{R}^{F_a \times 1}$, $b_\psi \in \mathbb{R}$, and $b_g \in \mathbb{R}^{F_a}$.

Another attention mechanism is introduced in the BD-LSTM when performing neural machine translation tasks. This attention allows paying different attention to parts of the encoded input information. Similarly, we would like to explore differences of the contributions that the encoded features and up-sampled features make to the segmentation task. This attention can be formulated by:

$$\beta_{i,t} = \frac{\exp\left(v_a^T \sigma_g(W_a[s_t, H_i])\right)}{\sum_{i'=1}^{n} \exp\left(v_a^T \sigma_g(W_a[s_t, H_i])\right)},$$

where $v_a$ and $W_a$ are transformation matrices, respectively. $\beta_{i,t}$ reflects quantitatively the importance of each hidden state of an input $x_i$ is to the corresponding output.

## 3    Experiments

In this section, we evaluate the effectiveness of our proposed method by conduct experiments on liver CT images. We adopt dice coefficient (DC) to measure the quality of segmentation, which is defined as follows:

$$DC(Y, Y') = \frac{2*(Y \cap Y')}{|Y| + |Y'|},$$

where $Y$ and $Y'$ denote the ground-truth and the predicted labels for each pixel in a given CT image, respectively.

### 3.1    Data Description

The data used in this study were collected from Pamela Youde Nethersole Eastern Hospital in Hong Kong, which contains 1 mm thin-sliced abdominal CT images of 571 patients. 72 patients were diagnosed with HCC and 499 patients had no HCC. CT images of each patient covered 4 phases, namely non-contrast, late-arterial, portal venous and delayed. The size of CT images is 512x512, while the slice number for each patient in each phase can vary, as summarized in Table 1 and Fig. 2.

Table 1 The statistical description of the number of slices over four phases.

| Phase | Number of slices (Range) |
| --- | --- |
| 1 (Non-contrast) | 145, 715 |
| 2 (Arterial) | 147, 401 |
| 3 (Portal venous) | 145, 903 |
| 4 (Delayed) | 148, 784 |

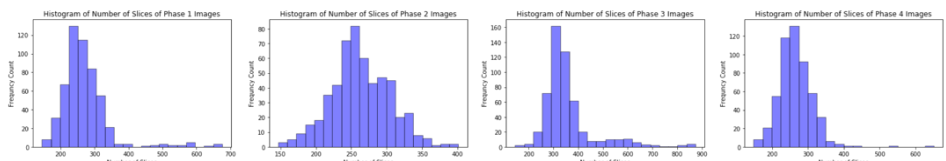

Fig. 2 The histogram of the slice numbers over four phases.

### 3.2 Quantitative Evaluation

In this part, we illustrate comparative results of our method DA-BDense-UNet against state-of-the-art segmentation models in Table 2. These counterparts include UNet [1], ResUNet [2], DenUNet [3]. Besides, we have integrated BDLSTM into these models to obtain BDLSTM-UNet, BDLSTM-ResUNet and BDLSTM-DenUNet, respectively. The reported best liver lesion segmentation result [4] has reached 0.8570, which is evaluated on the LITS data set. From Table 2, we observe ResUNet and DenUNet can achieve better segmentations than UNet does. The reason is that ResUNet adopts skip connections to regulate the identity information flow, which allows a deeper network capable of enhancing the representative power. Similarly, DenUNet improves the capability of network by densely connecting the preceding layers to all the succeeding layers. Meanwhile, the combined version of BDLSTM and UNet has achieved a better segmentation performance than UNet by about 1.2%. Similar observations are observed by comparing ResUNet and BDLSTM-ResUNet, DenUNet and BDLSTM-DenUNet, respectively. It indicates that BDLSTM can help to improve segmentation performance by exploring relationships between the encoded features and the up-sampled features. Our DA-BDense-UNet achieves a better result than BDLSTM-DenUNet by about 0.85%, due to two attention mechanisms introduced in the DenseUNet and BDLSTM. The attention gate in the DenseUNet allows focusing on pixelwise responses of salient regions of lesion candidates, and the attention in BDLSTM considers contribution differences between the encoded features and the up-sampled features to assign proper weights before conducting feature concatenation. As a result, DA-BDense-UNet delivers the best performance among the comparative methods.

Table 2 The comparison results obtained by counterparts and DA-BDense-UNet.

| Model | DC |
|---|---|
| UNet | 0.7846 |
| ResUNet | 0.8142 |
| DenUNet | 0.7989 |
| BDLSTM-UNet | 0.8066 |
| BDLSTN-ResUNet | 0.8290 |
| BDLSTM-DenUNet | 0.8435 |
| DA-BDense-UNet | 0.8520 |

In addition, we explore contributions of both the attention mechanisms in our method to the segmentation performance by conducting ablation study. Specifically, we remove the attention gate in the DenseUNet and the attention mechanism in BDLSTM and denote them by BDense-UNet-1 and BDense-UNet-2, respectively. The comparative results are shown in Table 3. Compared with BDLSTM-DenseUNet, we observe the attentions in the DenseUNet and BDLSTM enhance the performance by 0.47% and 0.63%, respectively. It indicates the effectiveness of the attention mechanisms. The attention in DenseUNet makes a larger contribution to performance improvement than the attention in BDLSTM, since the latter works on the foundation of the former which can accentuate the salient responses and depress irrelevant response. Besides, the combination of these two attentions delivers a better segmentation.

Table 3 The results of ablation study by removing the attention blocks in our method.

| Model | DC |
|---|---|
| BDLSTM-DenseUNet | 0.8435 |
| BDense-UNet-1 | 0.8482 |
| BDense-UNet-2 | 0.8498 |
| DA-BDense-UNet | 0.8520 |

### 3.3 Qualitive Evaluation

We show segmentation results obtained by our method together with the original CT image slices and the corresponding ground-truth in Fig. 3. From this figure, we notice that our method can identify the lesion position, achieving a satisfactory segmentation performance in the easy CT image slices, shown in the first and second rows of Fig. 3. Concerning difficult slices where there are three or more than lesions with different lesion size like the one in the third row, although our method is able to localize the lesions, the segmentation areas are smaller than those of the ground-truth ones.

Original Image     Ground-Truth     Prediction

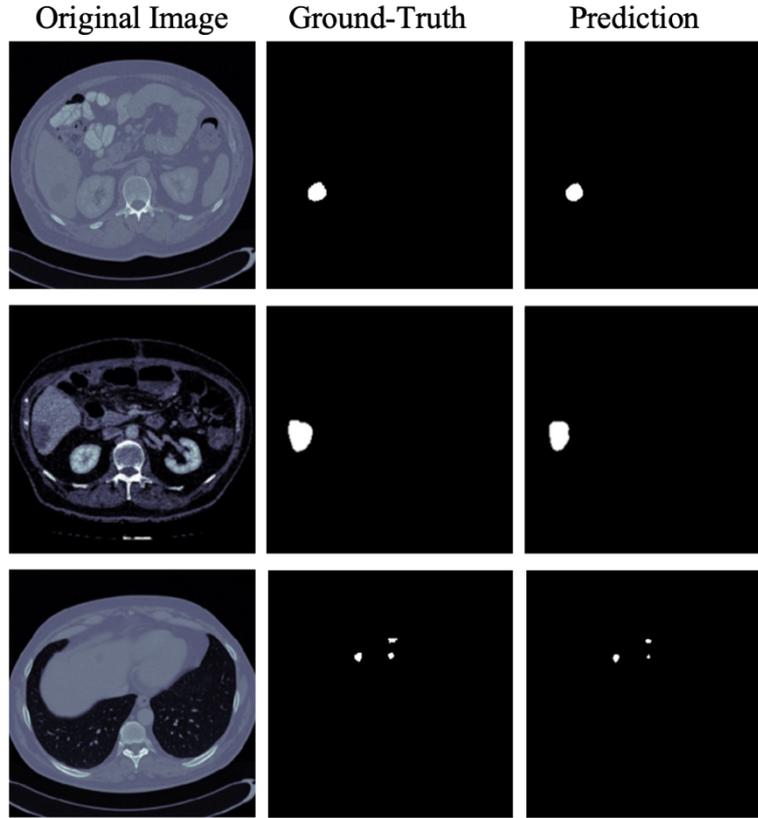

Figure 3 The examples of the original CT image slices, the corresponding ground-truth and the prediction results obtained by our method.

Additionally, we shall visualize the role of both attention mechanisms that play in the learning the feature maps, which contributes to the segmentation tasks. This can make a difference in interpreting how deep neural networks works in the segmentation. [The heat maps of channel-wise feature maps are expected to be available in the following weeks.]

## 4    Conclusion and Future work

In this work, we propose a new network scheme for automatic liver lesion segmentation, termed as DA-BDense-UNet. Firstly, we adopt densely connected convolutional networks as the backbones of UNet. Secondly, we introduce bidirectional long short term memory (BD-LSTM) to explore the correlations between the encoded features in the encoding path and the up-sampled features in the decoding path. Thirdly, we fuse attention mechanisms into convolutions and BD-LSTM. We conduct experiments on liver CT images collected from several hospitals. Experimental results verify the effectiveness of our method in terms of dice coefficient.

In the future, we plan to explore the feasibility of utilizing generative adversarial networks to synthesize diverse liver images. First, we can generate images with a specific kind of lesion by using CGAN, which mitigates the issues of data scarcity and class imbalance. Thus, it can improve performance of lesion classification. Second, we can utilize all the CT images and corresponding segmentation from one hospital and partial CT images and corresponding segmentations together with unlabeled CT images from another hospital to construct a semi-supervised segmentation scheme via domain adaptation, where GAN learns the distribution of lesion and generate images given the manual annotation of lesion.

Besides, we shall explore to transplant our designed classification and segmentation models trained on CT images to MR images by using the thought of domain adaptation and multimodal learning, since we consider easily the CT and MR images as the source and the target domains or two related but different modal data. In addition, it is possible to explore lesion segmentations over multiple organs simultaneously, such as the liver and lung.

## References

1. Ronneberger, Olaf, Philipp Fischer, Thomas Brox: U-net: convolutional networks for biomedical image segmentation. In International Conference on Medical Image Computing and Computer-Assisted Interventions, pp. 99–110, Springer, Cham, 2016.
2. M. P. Heinrich, Stille M., T. M. Buzug: Residual UNet convolutional neural network architecture for low-dose CT denoising. Current Directions in Biomedical Engineering, 4(1), pp. 297-300, 2018.
3. X. Li., H. Chen, X. Qi, Q. Dou, C.W. Fu, P. A. Heng: H-DenseUNet: hybrid densely connected UNet for liver and tumor segmentation from CT volumes. IEEE Transactions on Medical Imaging, 37 (12), pp. 2663–2674, 2016.
4. C. Sun, S. Guo, H. Zhang, J. Li, M. Chen, S. Ma, L. Jin, X. Liu, X. Li and X. Qian: Automatic segmentation of liver tumors from multiphase contrast-enhanced CT images based on FCNs, Artificial Intelligence in Medicine, 83, pp. 58-66, 2017.
5. F. Ouhmich, V. Agnus, V. Noblet, F. Heitz, P. Pessaux: Liver tissue segmentation in multiphase CT scans using cascaded convolutional neural networks. International Journal of Computer Assisted Radiology and Surgery, 14, pp. 1275-1284, 2019.
6. S. Lee, J.S. Bae, H. Kim, J.H. Kim, S. Yoon: Liver lesion detection from weakly-labeled multi-phase CT volumes with a grouped single shot multi-box detector, MICCAI, 2018.
7. Todoroki Y., X. Han, Iwamoto Y., L. Lin, H. Hu, Y. Chen: Detection of liver tumor candidates from CT images using deep convolutional neural networks. In International Conference on Innovation in Medicine and Healthcare, pp. 140-145, Springer, Cham, 2017.
8. D. Liang, L. Lin, H. Hu, Q. Zhang, Q. Chen, X. Han, Y. Chen: Combining convolutional and recurrent neural networks for classification of focal liver lesions in multi-phase CT images. In international conference on medical image computing and computer-assisted intervention, pp. 666-675, Springer, Cham, 2018.
9. C. Dong, Y. Chen, L. Lin, H. Hu, C. Jin, H. Yu, X. Han, T. Tomoko: Simultaneous segmentation of multiple organs using random walks. In journal of information processing, 24(2), pp. 320-329, 2016.

10. C. Dong, Y. Chen, A. Foruzan, L. Lin, X. Han, T. Tomoko, X. Wu, X. Gang, H. Jiang: Segmentation of liver and spleen based on computational anatomy models. In computers in biology and medicine, 67, pp. 146–160, 2015.

11. W. Liu, D. Anguelov, D. Erhan, C. Szegedy, S. Reed, C. Y. Fu, A. C. Berg: Ssd: Single shot multibox detector. In European conference on computer vision, pp. 21-37. Springer, Cham, 2016.

Acknowledgements:
Supported by the Innovation and Technology Fund (reference no: ITS/122/18FP), The Government of the Hong Kong Special Administrative Region.

Disclaimer:
"Any opinions, findings, conclusions or recommendations expressed in this material/event (or by members of the project team) do not reflect the views of the Government of the Hong Kong Special Administrative Region, the Innovation and Technology Commission or the Panel of Assessors for the Innovation and Technology Support Program of the Innovation and Technology Fund."